\documentstyle[multicol,aps,prb,epsfig]{revtex}
\begin{document}

\title{Dislocations and Bragg glasses in two dimensions}

\author{Pierre Le Doussal}
\address{CNRS-Laboratoire de Physique Th\'eorique de l'Ecole Normale 
Sup\'erieure \\
24 Rue Lhomond, Paris 75231, France}
\author{Thierry Giamarchi}
\address{Laboratoire de Physique des Solides, CNRS URA 02, Universit{\'e} 
Paris--Sud,
                   B{\^a}t. 510, 91405 Orsay, France }
\maketitle

\begin{abstract}
We discuss the question of the generation of topological defects
(dislocations) by quenched disorder in two dimensional periodic systems.
In a previous study [Phys. Rev. B {\bf 52} 1242 (1995)] 
we found that, contrarily to $d=3$,
unpaired dislocations appear in $d=2$ above a length scale
$\xi_D$, which we estimated. We extend this description to
include effects of freezing and pinning of dislocations at
low temperature. The resulting $\xi_D$ at low temperature
is found to be {\it larger} than our previous estimate,
which is recovered above a characteristic temperature.
The dependence of $\xi_D$ in the bare core energy 
of dislocations is a stretched exponential.
We stress that for all temperatures below melting
$\xi_D$ becomes arbitrarily large at weak disorder compared
to the translational order length $R_a \gg a$.
Thus there is a wide region of length scales, temperature and disorder
where Bragg glass like behavior should be observable.
\end{abstract}

\begin{multicols}{2}
An outstanding current problem in condensed matter
physics is to understand the physical properties of periodic
systems in presence of point impurities. This is important
for numerous experimental systems, such as vortices in type II
superconductors \cite{blatter_vortex_review,giamarchi_book_young},
Wigner crystallization of electrons
\cite{andrei_wigner_2d,perruchot_prl,li_wigner_conductivity_field,%
li_wigner_conductivity_density},
charge density waves \cite{gruner_revue_cdw} etc.. We have proposed 
\cite{giamarchi_vortex_long} that, contrarily
to previous claims,
such systems generically possess at weak disorder in three
dimensions a distinct thermodynamic glass phase, the Bragg glass, with 
perfect topological order (dislocation free) 
and quasi long range translational order. An immediate
consequence of this theory is that a phase transition away from 
the Bragg glass at which dislocations suddenly proliferate 
must occur upon increase of disorder (or field) which is determinant
for the phase diagram of type II superconductors
\cite{giamarchi_vortex_long,giamarchi_diagphas_prb}. Thus, whether
dislocations are generated or not by disorder has emerged as
a central question for these systems. In three dimensions
there is increasing theoretical
\cite{carpentier_bglass_layered,kierfeld_bglass_layered,fisher_bragg_proof}
, numerical 
\cite{gingras_dislocations_numerics,ryu_diagphas_numerics2,%
vanotterlo_bragg_numerics}
and experimental evidence
\cite{khaykovich_diagphas_bisco,%
khaykovich_bscco_irradiation,deligiannis_bglass_ybco,fuchs_bscco_surface}
that this topologically ordered Bragg glass phase exists.
By contrast, the strong disorder glassy state in three
dimensional superconductors
is still poorly known beyond the fact that it must contain
topological defects. It is yet unclear whether this state
is a distinct thermodynamic phase or simply a crossover 
from a pinned liquid. The initial proposal of a ``vortex glass'' phase
\cite{fisher_vortexglass_short,fisher_vortexglass_long} 
(claimed to contain topological defects)
now clearly cannot stand at weak disorder. It is even unclear,
in view of recent numerical 
\cite{bokil_young_vglass,kosterlitz_vortexglass}
and experimental data \cite{lopez_liquid_highfield}
whether such a phase, 
as described in Ref. 
\onlinecite{fisher_vortexglass_short,fisher_vortexglass_long},
exists at all in real (i.e with screening) three
dimensional superconductors.

In two dimensions it is easier to generate dislocations and
we have obtained \cite{giamarchi_vortex_long} that within the conventional
perturbative RG analysis unpaired dislocations appear beyond 
a length scale $\xi_D$ which we have estimated. A similar result 
was also obtained for the specific case of two dimensional 
triangular lattice \cite{carpentier_melting_prl,carpentier_ledou_triangco}.
Recent numerical studies
\cite{gingras_dislocations_numerics,zeng_fisher_nobglass2d} 
in $d=2$ support these predictions. Note however that
the question of the stability of the Bragg glass can only
be decided numerically by investigating the {\it weak disorder} region.
The RG flow used 
\cite{giamarchi_vortex_long}, although based on 
perturbation theory 
\cite{cardy_desordre_rg,rubinstein_shraiman_nelson}, 
already captured the delicate balance between the elastic
energy cost and the energy gain due to disorder. 
One striking consequence
was that although unpaired dislocations are generated, the scale
at which they appear, obtained as:
\begin{eqnarray}   \label{estimate0}
\xi_D \sim R_a e^{ c (\ln (R_a/a))^{1/2} }
\end{eqnarray}
can be made, at weak disorder $R_a \gg a$, 
arbitrarily large compared to the length
$R_a$ at which the displacements become of
the order of the lattice spacing.
\begin{figure}
\centerline{\epsfig{file=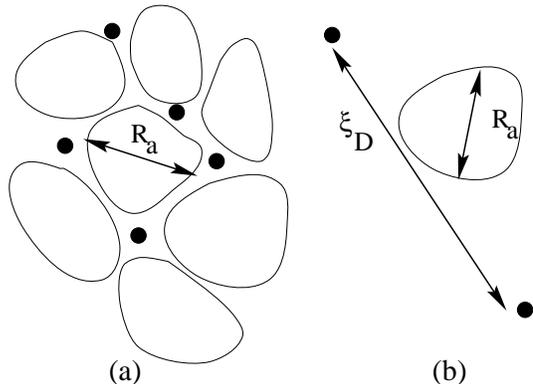,angle=-90,width=7cm}}
\caption{ \label{fig1} \narrowtext
(i) Conventional picture of a solid
broken up by disorder in domains of size $R_a$ 
with unpaired dislocations (black dots) appearing at the same
scale. (ii) correct picture for weak disorder:
the scale $\xi_D$ at which unpaired dislocations appear is 
larger than the scale $R_a$ at which translational order
starts decaying slowly.}
\end{figure} 
Below melting we thus obtained
that even in $d=2$ the naive picture of the system as 
breaking up in crystallites of size $R_a$ is still
incorrect as illustrated in Fig. \ref{fig1}.
Even if $d=2$ is the lower critical dimension
of the Bragg glass, since $\xi_D$ can be very large in practice
a wide regime of effective Bragg glass
behaviour should still be observable.
As pointed out in Ref. \onlinecite{carpentier_melting_prl}
where a theory predicting $\xi_D(T)$ near melting
was obtained, this may allow to understand why
experiments on two dimensional
superconductors \cite{ref_exp_2d} show 
a sharp change of behaviour with characteristics of
melting.

In this paper, we further examine the two dimensional
problem. Indeed there has been recent evidence that
at low temperature the conventional RG equations can {\it overestimate}
the importance of dislocations in related models
\cite{nattermann_xy_lowtemp,fertig_2d_disorder,%
korshunov_nattermann_diagphas,tang_xy_lowtemp,%
scheidl_xy_lowtemp,carpentier_xy_prl,carpentier_melting_prl,%
simus_randomphaseshift,scheild_lehnen}. Thus our
initial estimate (\ref{estimate0}) of $\xi_D$, which should be
accurate at intermediate temperatures, may become only
a lower bound on $\xi_D$ at low temperature. We give 
here a more accurate description of the low temperature
properties. This is of importance for applications
to experiments or simulations on classical systems,
often performed at low temperature but also for ground 
state properties of quantum disordered systems, such
as the two dimensional Wigner crystal 
\cite{giamarchi_columnar_variat,chitra_wigner_hall}.

In order to study dislocation problems in $d=2$ let us restrict
for simplicity to the single
component scalar XY model in a random field, defined in the 
continuum by the partition sum $Z = \int D\phi e^{-H/T}$ and
hamiltonian:
\begin{eqnarray} \label{2DXY}
H &=& \int d^2x\;\frac{J}{2 \pi} ({\bf \nabla} \phi(x) - {\bf
\eta}(x))^2 \\
& & - \zeta_1(x) \cos \phi(x) - \zeta_2(x) \sin \phi(x) \nonumber
\end{eqnarray}
with $\overline{ \eta_i(x) \eta_j(x') }=\pi \sigma \delta_{ij} \delta(x-x')$
and $\overline{ \zeta_i(x) \zeta_j(x') }= g \delta_{ij} \delta(x-x')$
are Gaussian white noises. Configurations with vortices 
(of integer charge $q_i$ at $x_i$) are described by
decomposing $\phi(x) = \phi_{SW}(x) + \phi_{V}(x)$
where $\phi_{SW}(x)$ is the smooth (spin wave) field and
$\phi_{V}(x)$ the vortex contribution
${\bf \nabla} \times {\bf \nabla} \phi_{V}(x) = \sum_i q_i 2 \pi \delta(x-x_i)$.
This model contains essential ingredients of a variety of elastic
disordered systems \cite{giamarchi_vortex_long} in $d=2$. For instance, in the case of
a lattice, $\phi = 2 \pi u/a$ can be thought of as a displacement field $u$
in units of lattice spacing $a$. Point like impurities produce
a random potential $V(x)$ which couples to the density and leads
to (\ref{2DXY}) with $g \sim \Delta_{K_0}/a^4$ proportional to the amplitude 
of the disorder with Fourier component close to $K_0=2 \pi/a$
(called the pinning disorder) and $\sigma \sim \Delta_{0}/4 \pi J^2 a^2$
proportional to the long wavelength disorder,
where $\Delta_q = \overline{V_q V_{-q}}$. Vortices in
the field $\phi$ correspond to dislocations in the lattice
and thus will generically be called ``dislocations'' in the 
following. These defects are characterized by a fugacity $y$, which 
in the bare model is related to the defect core energy $y = e^{- E_c/T}$.
Note that the long wavelength disorder $\sigma$ is always generated 
by coarse graining. The bare model (\ref{2DXY}) corresponds, 
for the lattice problem,
to the simpler case where the correlation length of the disorder $r_f$ is of
the order of the lattice spacing $a$. Thus in the model specifically
studied in this paper, the 
translational order correlation length \cite{giamarchi_vortex_long}
$R_a$ - such that relative displacements $u(R_a)-u(0) \sim a$,
is of the same order than the Larkin Ovchinnikov \cite{larkin_ovchinnikov_pinning}
pinning length $R_c$ - such that relative displacements $u(R_c)-u(0) \sim r_f$,
and we will implicitly equate them in the following. The situation $r_f \ll a$
(thus $R_c \ll R_a$) will be briefly mentioned at the end.

Let us summarize the RG analysis 
which led to the estimate in Ref.~\onlinecite{giamarchi_vortex_long} 
of the scale $\xi_D$ beyond
which unpaired dislocations appear in $d=2$ at weak disorder.
The RG equations for the fugacity of dislocations,
the disorder and the stiffness were derived by Cardy and Ostlund (CO)
\cite{cardy_desordre_rg,rubinstein_shraiman_nelson}.
The fugacity of the vortices $y$ satisfies to lowest order in $y$:
\begin{equation}\label{fugacity1}
\frac{dy}{dl} = (2 - \frac{J}{T} + \frac{\sigma J^2}{T^2} ) ~~y
\end{equation}
with $T_m=J/2$ is the pure system Kosterlitz-Thouless 
transition temperature and $l=\ln L$ is the logarithmic scale.
The pinning disorder renormalizes as:
\begin{equation} \label{fugacity2}
\frac{d g}{dl} = 2 \tau g - B g^2 
\end{equation}
up to $O(g^3)$ terms, with $\tau=1 - \frac{T}{T_g}$, 
$T_g=4 J=8 T_m$ and $B$ a nonuniversal 
constant. If dislocations are excluded by hand (setting $y=0$)
there is a transition at $T=T_g$ between a high temperature
phase ($T> T_g$) where the disorder is irrelevant 
and a low temperature glass phase ($T < T_g$).
For $T<T_g$ there is a line of fixed points $g=g^*$ which describes
a 2d Bragg glass phase where displacements grow 
as $u \sim \ln x$ and beyond which translational order
slowly decays. This asymptotic behaviour is reached at
the length $R_a \sim a (\frac{g^*}{g_0})^{\frac{1}{2 \tau}}$
(where $g_0$ is the bare value)
which for weak disorder is $R_a \gg a$. 
The global structure of the RG
(together with functional RG extensions) suggests that the fixed line 
continues down to $T=0$ and that correlations behave as 
$\overline{\langle(\phi(x) - \phi(0))^2\rangle} = C(T) 
\ln^2(x/R_a)$ with a $T$ dependent prefactor \cite{toner_log_2}.
It is perturbatively controlled only near $T_g$, where $g^*$ is small,
and the correct universal prefactor \cite{footnote1} $C(T)=2 \tau^2$ was derived 
\cite{carpentier_ledou_triangco}
to lowest order in $\tau$. Numerical simulations
have measured \cite{footnote2} $C(T)$
near $T_g$ \cite{marinari_simu_logcarre} and at $T=0$ \cite{zeng_logcarre,%
rieger_blasum_logcarre,zeng_fisher_nobglass2d}

When dislocations are allowed in the model,
one finds that
they are perturbatively relevant above $T_m$ leading to
a liquid. Below $T_m$ however, since the 
bare value of $\sigma$ is small for weak disorder, one could
naively conclude from (\ref{fugacity1}) that dislocations are 
suppressed at low enough temperature.
As was discussed in Ref.~\onlinecite{giamarchi_vortex_long} 
this is not the case and
in fact the 2d Bragg glass fixed point
is {\it unstable} to dislocations. Such an instability
is a peculiar feature of $d=2$ and does not occur
in $d=3$ where the Bragg glass phase is {\it stable} with respect
to dislocations at weak disorder. The instability occurs
because in $d=2$ the long wavelength disorder $\sigma$ 
is also renormalized:
\begin{equation} \label{fugacitysig}
\frac{d \sigma}{dl} = A g^2
\end{equation}
to lowest order (with $A=B^2$ at $T=T_g$)
and thus grows unboundedly with the scale as $\sigma(l) \sim \sigma^* + 2 C(T)
\ln(L/R_a)$ for $L>R_a$, with $2 C(T)= A {g^*}^2=4 \tau^2$ near $T_g$.
As can be seen from (\ref{fugacity1})
$y(l)$ starts increasing beyond a certain length scale
and eventually becomes of order $y(l_D) \sim 1$ 
at a scale $l_D= \ln (\xi_D/a)$.
At that scale unpaired dislocations dominate the behaviour
and translational order is exponentially destroyed beyond $\xi_D$.
\begin{figure}
\centerline{\epsfig{file=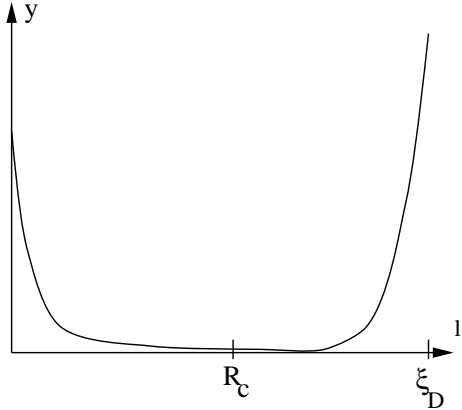,angle=-90,width=6cm}}
\caption{\label{fig2} Dependence of the dislocation
fugacity $y$ in the length scale for weak disorder. It is first strongly
renormalized downwards before it eventually shoots up
again and reaches values of order unity at $\xi_D \gg R_a$.}
\end{figure}
However,
$\xi_D$ can be very large
and in particular much larger than $R_a$.
This is because for weak disorder and $T < T_m$ the fugacity of dislocations
is first strongly renormalized downwards for scales
smaller than $R_a$ (see (\ref{fugacity1})),
as shown in Figure \ref{fig2}.
Integrating the
flow led us to the estimate \cite{giamarchi_vortex_long}:
\begin{eqnarray} \label{estimate}
\xi_D \sim R_a e^{ c \sqrt{ 2 (\frac{T_m(\sigma)}{T} -1) \ln (R_a/a) } }
\end{eqnarray}
where $T_m(\sigma)$ is the boundary of the XY phase in the absence of
pinning disorder ($g=0$) as given by the CO equations
\cite{rubinstein_shraiman_nelson} (see Fig. \ref{fig4} below).
The case of triangular lattices
in presence of disorder was studied in more details recently
using a $N=2$ component model, and a generalization of
both the above CO equations 
(\ref{fugacity1},\ref{fugacity2},\ref{fugacitysig}) 
and the KTNHY equations
(which describe the fusion of pure crystals) was obtained
\cite{carpentier_melting_prl,carpentier_xy_long}.
It confirms that a similar estimate as
(\ref{estimate}) holds for lattices and yields a complementary
formula for $\xi_D$ around the pure crystal melting 
transition.

To obtain (\ref{estimate}) the only assumption
made in Ref. \onlinecite{giamarchi_vortex_long} and supported
by the structure of the RG flow, was that beyond the
scale $R_a$ pinning disorder $g$ has reached a fixed point
and that $\sigma(l)$ grows as $l=\ln L$. This is equivalent 
to use beyond $R_a$ the effective ``random stress'' model
which reads in the continuum:
\begin{eqnarray} \label{2DXYbis}
H &=& \int d^2x\;\frac{J}{2 \pi} ({\bf \nabla} \phi(x) - {\bf \eta}(x))^2 
\end{eqnarray}
to which one must add vortices as discussed above. The disorder 
$\overline{ \eta_i(q) \eta_j(-q) }=\pi \sigma(q) \delta_{ij}$,
with $\sigma(q) \sim \ln(1/q)$, 
now depends on the scale in a logarithmic way as
$\sigma(l) \sim \sigma^* + 2 C(T) \ln(L/R_a)$, where
$C(T)$ is the amplitude defined above. A recent numerical 
work \cite{zeng_fisher_nobglass2d}
comes as additional support that this assumption is indeed
valid. When $\sigma(l)=\sigma$ is 
scale independent, this model reduces to the ``random phase shift
model'' \cite{rubinstein_shraiman_nelson}.
The advantage of model (\ref{2DXYbis})
is that it can be treated by RG or by qualitative
arguments even in presence of dislocations.

Let us now reexamine (\ref{2DXYbis}) at low temperature. 
Because of the reentrance of the disordered phase present
in the CO equations \cite{rubinstein_shraiman_nelson}
an extrapolation of (\ref{estimate}) would lead to a small
$\xi_D$ at low temperature (below $T^{-}_m(\sigma)$
in Fig. \ref{fig4}). However, the original CO
equations tend to {\it overestimate} the effect of dislocations 
as they neglect non thermalization and pinning of dislocations
\cite{giamarchi_vortex_long}. Indeed,
recent reexamination of the phase diagram of the
random phase shift model (\ref{2DXYbis}) has shown that 
the reentrant disordered phase 
\cite{rubinstein_shraiman_nelson,cardy_desordre_rg}
disappears when these effects are taken into account
\cite{nattermann_xy_lowtemp}. Although in the high temperature region 
$T>T_m/2$ the RG equation (\ref{fugacity1}) was found to be correct, 
new physics arises below the line $T < T^*(\sigma)$.
In order to reexamine our previous estimate (\ref{estimate}) for $\xi_D$
at low temperature, we extend to the present case 
(where $\sigma(l)$ depends logarithmically on $L$)
the modified RG analysis recently developed for the random phase
shift model \cite{nattermann_xy_lowtemp,korshunov_nattermann_diagphas,%
tang_xy_lowtemp,scheidl_xy_lowtemp,carpentier_xy_prl,%
carpentier_xy_long}.

To study the relevance of dislocations it is first useful 
to consider a single dislocation or dipole at $T=0$.
The simplest energy argument, presented first in 
Ref. \onlinecite{nattermann_xy_lowtemp,korshunov_nattermann_energy},
estimates when it is favorable to place one vortex or a dipole 
in the system of size $L$. This vortex sees a 2d random potential $V(x)$
with logarithmic correlations, and one must thus estimate the minimum 
energy $E_{min}$ of this random potential.
A reasonable approximation is obtained
by neglecting correlations but keeping the correct local
variance as \cite{korshunov_nattermann_energy}:

\begin{eqnarray}  \label{argument}
\frac{1}{L^2} \sim \int_{- \infty}^{E_{min}} \frac{dV}{\sqrt{4 \pi \sigma J^2 \ln L}}
\exp(- \frac{V^2}{4 \sigma J^2 \ln L} ) 
\end{eqnarray}

This estimate leads to $E_{min} \sim - \sqrt{ 8 \sigma} J \ln L$. 
The more accurate methods which are now available to
take correlations into account \cite{castillo97,carpentier_xy_prl}
confirm that the prefactor is exact, and estimate 
the (large) corrections to scaling. The energy gain by disorder thus
overcomes the elastic cost $E_{el} = J \ln L$ for $\sigma > \sigma_c = 1/8$.
Below this value an XY phase, dislocation free at large scale,
exists contrarily to the earlier conclusion \cite{rubinstein_shraiman_nelson}
based on (\ref{fugacity1}). Thus these low temperature
effects strongly reduce the relevance of dislocations.
In the random phase shift model, the topologically 
ordered Bragg glass phase is thermodynamically stable 
at low temperatures.

One can thus expect a similar {\it reduction} of the 
importance of dislocations in the presence of
pinning disorder, i.e for the random stress
model with $\sigma(l) \sim \ln L$ compared to
our previous estimates based on the CO equations.
A straightforward modification of the above argument 
(\ref{argument}), replacing $\sigma$ by 
$\sigma(l)$ in (\ref{argument}) leads
to $E_{min} \sim \sqrt{ 8 \sigma(l) } J \ln L  \sim (\ln L)^{3/2}$.
The energy gain due to disorder now always overcomes the
elastic cost $J \ln L$ at large scale and dislocations are
always generated in model (\ref{2DXYbis}). This modification
of the argument of Ref. \onlinecite{korshunov_nattermann_energy} thus
allows to recover in a very simple 
way the $(\ln L)^{3/2}$ estimate \cite{zeng_fisher_nobglass2d}
for the optimal energy of a dislocation pair.

However these types of energy arguments or 
the analysis \cite{zeng_fisher_nobglass2d} for a {\it single} dislocation
(or dipole) does not by itself allow to compute
the length scale $\xi_D$ at which vortices (dislocations) destroy XY
(positional) order. Indeed, to destroy the order exponentially one needs a
{\it finite density} of defects.
In particular it would be {\it incorrect} to
identify $\xi_D$ as the length scale at which the disorder
energy becomes of the order of the elastic energy. This is 
clear for instance when looking at the KT transition were
elastic energy is $J \ln L$ and entropy is $2 T \ln L$.
When $T > T_m =J/2$ balancing only these two terms would incorrectly
predict unpaired dislocations
of size $a$ near $T_m$, when in reality they occur at a much
larger length scale $\xi_m \sim a e^{-c (T-T_m)^{-1/2}}$. To get the
correct result one must take into account both the defect
core energy $E_c$ and the screening by smaller dipoles. The same
effects {\it must} be also taken into account for the random
stress model in order to quantify the importance of dislocations.

\begin{figure}
\centerline{\epsfig{file=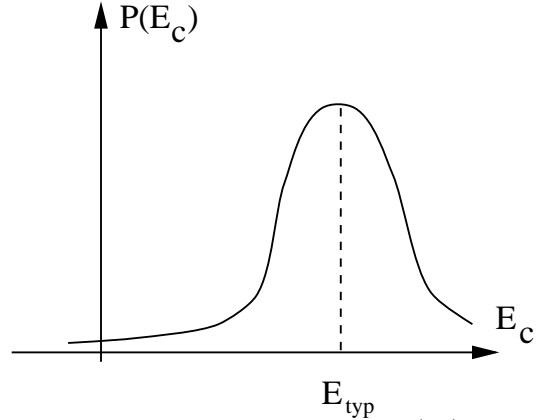,angle=-90,width=7cm}}
\caption{ \label{fig3} scale dependent distribution $P_l(E_c)$
of the effective core energy as
discussed in Ref. {\protect \onlinecite{carpentier_xy_prl}}. The bulk 
of the distribution is centered around a typical value $E_{typ}(l)$ but 
broad tails develop at low temperature}
\end{figure}

We now estimate the length scale $\xi_D$ at low temperature
using a RG analysis
for the random stress model (\ref{2DXYbis}). 
In a recent work \cite{carpentier_xy_prl} it was shown that 
to describe the correct physics in the model 
with constant $\sigma(l)=\sigma$, one must follow the full
distribution of dislocation core energies $P_l(E_c)$, found to satisfy 
a non linear RG equation. At $T=0$, this non linear RG equation solves
the problem of energy minimization iteratively on
successive logarithmic scales. Its solution $P_l(E_c)$, 
develops broad tails at low temperature, as illustrated in 
Fig. \ref{fig3}. Very schematically, while the center 
of the distribution is biased towards the right as $\sim J \ln L$, the
width grows as $\sim \sigma J^2 \ln L$. For large enough
$\sigma$ ($\sigma >1/8$) the width ``wins'' and the probability
of finding a site with negative effective core energy
eventually increases with the scale. This analysis 
can be extended in an accurate manner to the
present case, by studying the non linear RG equation in presence
of a scale dependent $\sigma(l) \sim l$.

For our present purpose it is enough to use the following
simplified version where $P_l(E_c)$ at scale $l$ is given
by:
\begin{eqnarray}  \label{gaussian}
P_l(E_c) \sim \exp( 2 l - 
\frac{(E_c - J l - E_c^0)^2}{4 J^2 \int_0^l \sigma(l') dl' + c} )
\end{eqnarray}
This amounts to study the solution of the linearized version of
the RG equation \cite{carpentier_xy_prl} which
should be a good approximation in the small $E_c$ far tail 
of the distribution of $P_l(E_c)$ needed here. 
The distance between unpaired dislocations corresponds to the length scale
$l = l_D = \ln (\xi_D/a)$ at which $P_l(0) \sim 1$. Physically 
this corresponds, in the renormalized model, to putting one dislocation
per site. The above form, for constant $\sigma$, clearly 
yields a dislocation free phase ($l_D=+\infty$) for $\sigma<1/8$
and proliferation of dislocations for $\sigma>1/8$. In that case
differentiating $P_l(0)$ from (\ref{gaussian}) 
yields back the RG equation \cite{nattermann_xy_lowtemp} for
the ``effective dipole fugacity'' $y^2$ as:
\begin{eqnarray}  \label{thomas}
\frac{dy}{dl} = (2 - \frac{1}{4 \sigma})~y \qquad T<T^*=2 \sigma J
\end{eqnarray}
identified as $y^2 \sim P_l(0)^2$ below the line $T=T*=2 \sigma J$ 
where the freezing 
does occur (see Figure \ref{fig3}).
In the present case the above equation (\ref{gaussian}) 
is complemented by the equation for $\sigma(l)$ which arises 
as the solution of (\ref{fugacity2},\ref{fugacitysig}).
Denoting $R_a=a e^{l_a}$, and using the simplified
form $g(l)= g^* e^{2 \tau (l-l_a)}$ for $l<l_a$
and $g(l)= g^*$ for $l > l_a$, we obtain:
\begin{eqnarray}
&& \sigma(l) = \sigma_0 + \frac{A {g^*}^2}{4 \tau} (e^{4 \tau (l-l_a)} -
e^{- 4 \tau l_a} ) \qquad l<l_a \nonumber \\
&& \sigma(l)  \approx \sigma^* + A {g^*}^2 (l-l_a)  \qquad l > l_a
\label{sig}
\end{eqnarray}
with $\sigma^* = \frac{A {g^*}^2}{4 \tau}$. This is valid
near $T_g$, but can be extended everywhere by 
replacing $A {g^*}^2$ by $2 C(T)$ in (\ref{sig}).
Also, in the range of length
scales needed here, and far from the critical crossover region 
$\sigma_0=1/8$, we can neglect the renormalization of
$\sigma$ and $J$ by dislocations.
Using (\ref{sig}) , (\ref{gaussian}) leads to:
\begin{eqnarray}  \label{xilowt}
\xi_D \sim R_a e^{ c \sqrt{(\frac{1}{8} -  \sigma_0) \ln (R_a/a)} }
\end{eqnarray}
with $c \approx 1/\sqrt{C(T)}$.
This expression holds in the low temperature 
region $T < T^* = 2 \sigma J$ and for weak disorder
$R_a \gg a$. Interestingly, it does have a form 
very similar to (\ref{estimate}) which is valid in the high temperature region
$T^* < T < T_m(\sigma)$ in Fig. \ref{fig4}, but for the fact that
now the disorder $\sigma_0$ plays a role analogous to temperature.
The expression (\ref{xilowt}) smoothly connects with our previous
estimate (\ref{estimate}) at higher temperature. It would be
interesting to check the predictions (\ref{estimate},\ref{xilowt}) 
in numerical simulations. A previous numerical work 
\cite{gingras_dislocations_numerics} on the random XY model
close to $T_m$ is indeed consistent with $\xi_D$ larger
than $R_a$ although no attempts were made to compare the
results with (\ref{estimate}). Finally, very near $\sigma_0=1/8$
we expect that $\xi_D$ will take a critical crossover scaling
form as is the case \cite{carpentier_melting_prl} near $T=T_m$. 
\begin{figure}
\centerline{\epsfig{file=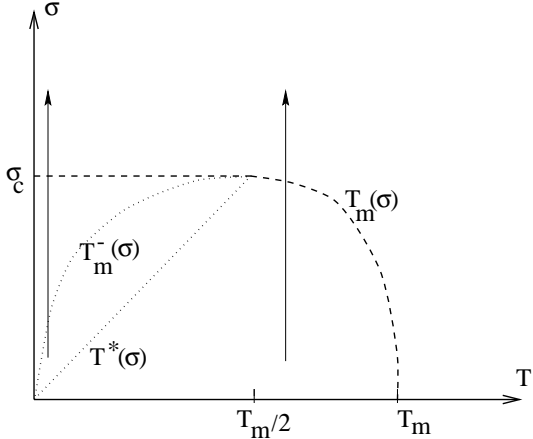,angle=-90,width=7cm}}
\caption{\label{fig4}
Different regions of the temperature $T$, disorder $\sigma$
phase diagram. The dashed lines represent true XY phase boundaries 
($T_m(\sigma)$ and $\sigma=\sigma_c=1/8$)
for the model without pinning disorder $g=0$. The dotted lines
are the old (incorrect) XY boundary $T^{-}_m(\sigma)$
obtained via CO-RG, and the new freezing line $T^*(\sigma)$.
In presence of 
pinning disorder $g>0$, $\sigma(l)$ grows with the scale and
eventually dislocations appear. The arrowed lines represent
the RG trajectories in the two cases (low and intermediate temperature)
discussed in the text}
\end{figure}

Note that it would be tempting to replace
directly $\sigma$ by $\sigma(l)$ in (\ref{thomas})
and integrate the flow. This is incorrect, 
as can be seen by writing
the RG equation for $P_l(0)$ using (\ref{gaussian}).
This is a manifestation of the fact that one must here 
follow a distribution which cannot, at low temperature, 
be parametrized by
a single fugacity-like parameter $y$.

In numerical simulations it 
may be easier to study intermediate disorder but
change artificially the bare core energy $E_c^0$. 
Keeping the core energy dependence in the previous
analysis of (\ref{gaussian}) leads to the following
equation for $l_D = \ln(\xi_D/a)$:
\begin{eqnarray}
0 = 2 l_D - \frac{(J l_D + E_c^0)^2}{4 J^2 C(T) l_D^2}
\end{eqnarray}
In the limit of large $E_c^0$ this gives:
\begin{eqnarray}  \label{deuxtiers}
\xi_D \sim a ~~ e^{ (\frac{E_c^0}{E^*})^{\frac{2}{3}} }
\end{eqnarray}
with $E^* \approx 2 \sqrt{2 C(T)} J $. 
It is noteworthy that a very recent numerical
work \cite{zeng_leath_nobglass2d}
finds a similar law with an exponent $\sim 0.7$.
The above result can be written $E_c^0 \sim (\ln \xi_D)^{3/2}$
in agreement with the general scaling of optimal disorder energies
in this problem.

Let us close with an overall perspective on the situation
in $d=2$. The above analysis of the flow of pinning disorder
can be improved, e.g. allowing for more general RG flow
structure or using functional RG techniques appropriate
to zero temperature (where the model (\ref{2DXY}) develops
higher harmonics and nonanalyticity). This, we expect
should not change the results obtained here, up to
numerical prefactors, as long as the approximation 
introduced in Ref. \onlinecite{giamarchi_vortex_long}
by the random stress model holds at large scale. 
This model allows to treat some of the conventionally
non perturbative effects related to
dislocation freezing by long wavelength disorder.
Proving its validity everywhere, or going beyond it,
is difficult analytically, and challenging numerically
because of large corrections to scaling.
Thus it still cannot be excluded that more surprises 
may be in store when further non perturbative 
effects (e.g. the effect of pinning disorder $g$ on dislocations)
are taken into account. The case $r_f \ll a$ can also
be studied. An intermediate ``random manifold regime'' then exists between
$R_c < L < R_a \sim R_c (a/r_f)^\alpha$ with $\alpha \sim 3$.
A (functional) RG analysis can be sketched.
The RG equation for $\sigma$ becomes $\partial_l \sigma \sim \sum_K K^2 g_K$
to lowest order, while the $g_K$ reach a crossover functional form
beyond $L=R_c$ and another, asymptotic one, beyond $L=R_a$.
Although $\sigma(l)$ starts growing beyond $R_c$, we find that
it remains
small until $R_a$. This indicates that the above results 
should still hold.

To conclude we have analyzed the question of the generation
of dislocations by disorder in $d=2$. In an ealier study 
\cite{giamarchi_vortex_long}
based on a random stress model approximation of the
Cardy Ostlund equations we had found that contrarily to
$d=3$, in $d=2$ unpaired dislocations appear beyond 
a length scale $\xi_D$. Recent numerical simulations
\cite{gingras_dislocations_numerics,zeng_fisher_nobglass2d} 
have reached a similar conclusion. At low temperature
we have improved our estimate for $\xi_D$ taking into
account freezing of dislocations by long wavelength
disorder. Taking into account these effects increases
$\xi_D$ compared to our previous estimates at low
temperature and smoothly connects to it at higher temperature.
Thus the range of length scales where the system
behaves {\it effectively} as a Bragg glass is wider than
previously expected. We also computed the bare core energy
dependence of $\xi_D$ which exhibits stretched exponential
behaviour. It would be interesting to further
explore numerically these systems, particularly at
weak disorder.

\acknowledgements

We thank David Carpentier for useful discussions.
T.G. acknowledge support and hospitality from
I.T.P. (Santa Barbara)
where part of this work was completed.
This research was supported in part by the
National Science Foundation under grant 
PHY94-07194. 


\begin{thebibliography}{10}

\bibitem{blatter_vortex_review}
G. Blatter {\it et~al.}, Rev. Mod. Phys. {\bf 66},  1125  (1994).

\bibitem{giamarchi_book_young}
T. Giamarchi and P. {Le Doussal},  in {\em Statics and dynamics of disordered
  elastic systems}, edited by A.~P. Young (World Scientific, Singapore, 1998),
  p.\ 321, cond-mat/9705096.

\bibitem{andrei_wigner_2d}
E.~Y. Andrei and {al.}, Phys. Rev. Lett. {\bf 60},  2765  (1988).

\bibitem{perruchot_prl}
F. Perruchot et al. submitted to Phys. Rev. Lett. 1997.

\bibitem{li_wigner_conductivity_field}
C.~C. Li and al., Phys. Rev. Lett. {\bf 79},  1353  (1998).

\bibitem{li_wigner_conductivity_density}
C.~C. Li and al, 1998, cond-mat/9810066.

\bibitem{gruner_revue_cdw}
G. Gr{\"u}ner, Rev. Mod. Phys. {\bf 60},  1129  (1988).

\bibitem{giamarchi_vortex_long}
T. Giamarchi and P. {Le Doussal}, Phys. Rev. B {\bf 52},  1242  (1995).

\bibitem{giamarchi_diagphas_prb}
T. Giamarchi and P. {Le Doussal}, Phys. Rev. B {\bf 55},  6577  (1997).

\bibitem{carpentier_bglass_layered}
D. Carpentier, P. {Le Doussal}, and T. Giamarchi, Europhys. Lett. {\bf 35},
  379  (1996).

\bibitem{kierfeld_bglass_layered}
J. Kierfeld, T. Nattermann, and T. Hwa, Phys. Rev. B {\bf 55},  626  (1997).

\bibitem{fisher_bragg_proof}
D.~S. Fisher, Phys. Rev. Lett. {\bf 78},  1964  (1997).

\bibitem{gingras_dislocations_numerics}
M.~J.~P. Gingras and D.~A. Huse, Phys. Rev. B {\bf 53},  15193  (1996).

\bibitem{ryu_diagphas_numerics2}
S. Ryu, A. Kapitulnik, and S. Doniach, Phys. Rev. Lett. {\bf 77},  2300
  (1996).

\bibitem{vanotterlo_bragg_numerics}
A.~V. Otterlo, R. Scalettar, and G. Zimanyi, Phys. Rev. Lett. {\bf 81},  1497
  (1998).

\bibitem{khaykovich_diagphas_bisco}
B. Khaykovich and al., Phys. Rev. Lett. {\bf 76},  2555  (1996).

\bibitem{khaykovich_bscco_irradiation}
B. Khaykovich and al., Phys. Rev. B {\bf 56},  R517  (1997).

\bibitem{deligiannis_bglass_ybco}
K. Deligiannis and al., Phys. Rev. Lett. {\bf 79},  2121  (1997).

\bibitem{fuchs_bscco_surface}
D. Fuchs and al, 1998, cond-mat/9807016.

\bibitem{fisher_vortexglass_short}
M.~P.~A. Fisher, Phys. Rev. Lett. {\bf 62},  1415  (1989).

\bibitem{fisher_vortexglass_long}
D.~S. Fisher, M.~P.~A. Fisher, and D.~A. Huse, Phys. Rev. B {\bf 43},  130
  (1990).

\bibitem{bokil_young_vglass}
H.~S. Bokil and A.~P. Young, Phys. Rev. Lett. {\bf 74},  3021  (1995).

\bibitem{kosterlitz_vortexglass}
J. M. Kosterlitz and N. Akino, cond-mat/9806346.

\bibitem{lopez_liquid_highfield}
D. Lopez and al., Phys. Rev. Lett. {\bf 80},  1070  (1998).

\bibitem{carpentier_melting_prl}
D. Carpentier and P. {Le Doussal}, Phys. Rev. Lett. {\bf 81},  1881  (1998).

\bibitem{carpentier_ledou_triangco}
D. Carpentier and P. {Le Doussal}, Phys. Rev. B {\bf 55},  12128  (1997).

\bibitem{zeng_fisher_nobglass2d}
C. Zeng, P.~L. Leath, and D.~S. Fisher, cond-mat/9807281.

\bibitem{cardy_desordre_rg}
J.~L. Cardy and S. Ostlund, Phys. Rev. B {\bf 25},  6899  (1982).

\bibitem{rubinstein_shraiman_nelson}
M. Rubinstein, B. Shraiman, and D. Nelson, Phys. Rev. B {\bf 27},  1800
  (1983).

\bibitem{ref_exp_2d}
R. W\"ordenveber, P. H. Kes and C.C. Tsuei Phys. Rev. B {\bf 33} 3172 (1986),
  P. Berghuis et al. Phys. Rev. Lett. 65 2583 (1990), A. Yazdani Stanford PhD
  thesis (1994) and references therein.

\bibitem{nattermann_xy_lowtemp}
T. Nattermann, S. Scheidl, S.~E. Korshunov, and M.~S. Li, J. de Phys. I {\bf
  5},  565  (1995).

\bibitem{fertig_2d_disorder}
M. Cha and H.~A. Fertig, Phys. Rev. Lett. {\bf 74},  4867  (1995).

\bibitem{korshunov_nattermann_diagphas}
S.~E. Korshunov and T. Nattermann, Phys. Rev. B {\bf 53},  2746  (1996).

\bibitem{tang_xy_lowtemp}
L.~H. Tang, Phys. Rev. B {\bf 54},  3350  (1996).

\bibitem{scheidl_xy_lowtemp}
S. Scheidl, Phys. Rev. B {\bf 55},  457  (1997).

\bibitem{carpentier_xy_prl}
D. Carpentier and P. {Le Doussal}, Phys. Rev. Lett. {\bf 81},  2558  (1998).

\bibitem{simus_randomphaseshift}
J. Maucourt, D.R. Grempel cond-mat/9703109, J.M. Kosterlitz, M.V. Simkin, Phys.
  Rev. Lett. 79 (1997) 1098.

\bibitem{scheild_lehnen}
S. Scheidl and M. Lehnen, cond-mat/9802111.

\bibitem{giamarchi_columnar_variat}
T. Giamarchi and P. {Le Doussal}, Phys. Rev. B {\bf 53},  15206  (1996).

\bibitem{chitra_wigner_hall}
R. Chitra, T. Giamarchi, and P. {Le Doussal}, Phys. Rev. Lett. {\bf 80},  3827
  (1998).

\bibitem{larkin_ovchinnikov_pinning}
A.~I. Larkin and Y.~N. Ovchinnikov, J. Low Temp. Phys {\bf 34},  409  (1979).

\bibitem{toner_log_2}
J. Toner and D.~P. DiVincenzo, Phys. Rev. B {\bf 41},  632  (1990).

\bibitem{footnote1}
an unambiguous definition of $T$ is given by the large distance prefactor of
  the averaged connected correlations.

\bibitem{footnote2}
From what we extracted the measured $C(T)/\tau^2$ was about 0.57 in Ref.
  \onlinecite{rieger_blasum_logcarre}, 0.54 in Ref.
  \onlinecite{zeng_fisher_nobglass2d} both at $T=0$ and about 1.3 in Ref.
  \onlinecite{marinari_simu_logcarre} at $T>0$.

\bibitem{marinari_simu_logcarre}
E. Marinari, R. Monasson, and J.J. Ruiz-Lorenzo, J. Phys. A {\bf 28} 3975
  (1995).

\bibitem{zeng_logcarre}
C. Zeng, A.~A. Middleton, and Y. Shapir, Phys. Rev. Lett. {\bf 77},  3204
  (1996).

\bibitem{rieger_blasum_logcarre}
H. Rieger, U. Blasum cond-mat/9608136.

\bibitem{carpentier_xy_long}
D. Carpentier and P. {Le Doussal}, in preparation.

\bibitem{korshunov_nattermann_energy}
S.~E. Korshunov and T. Nattermann, Physica B {\bf 222},  280  (1996).

\bibitem{castillo97}
H. Castillo and {\it al.}, Phys. Rev. B {\bf 56},  10668  (1997).

\bibitem{zeng_leath_nobglass2d}
C. Zeng and P.~L. Leath, cond-mat/9810154.

\end{thebibliography}

\end{multicols}
\end{document}